\documentclass[sigconf, nonacm]{acmart}

\newcommand\vldbyear{2026}
\newcommand\vldbworkshop{DASHSys: Systems for Data-centric Agents with Human-in-the-loop}
\newcommand\vldbauthors{\authors}
\newcommand\vldbtitle{\shorttitle} 
\newcommand\vldbavailabilityurl{https://github.com/BauplanLabs/git_for_data}
\newcommand\vldbpagestyle{empty} 

\usepackage{array}
\usepackage{listings}
\floatstyle{ruled}
\newfloat{listing}{tb}{lst}{}
\floatname{listing}{Listing}

\begin{document}
\title{\textit{GitLake}: Git-for-data for the agentic lakehouse}
\author{Weiming Sheng}
\authornote{All authors contributed equally and are listed \texttt{ORDER BY AGE ASC}. JT is the corresponding author and PI on the project:  \url{mailto:jacopo.tagliabue@bauplanlabs.com}.}
\affiliation{%
  \institution{Columbia University}
  \country{USA}
}

\author{Jinlang Wang}
\authornotemark[1]
\affiliation{%
  \institution{University of Wisconsin-Madison}
  \country{USA}
}

\author{Manuel Barros}
\authornotemark[1]
\affiliation{%
  \institution{Carnegie Mellon University}
  \country{USA}
}

\author{Aldrin Montana}
\authornotemark[1]
\affiliation{%
 \institution{Bauplan Labs}
 \country{USA}
}

\author{Jacopo Tagliabue}
\authornotemark[1]
\affiliation{%
 \institution{Bauplan Labs}
 \country{USA}
}

\author{Luca Bigon}
\authornotemark[1]
\affiliation{%
  \institution{Bauplan Labs}
  \country{USA}
}

\begin{abstract}
We present \texttt{GitLake}, a Git-for-data design for an agent-first lakehouse. The system lifts single-table Iceberg snapshots into lakehouse-wide commits, branches, and merges, letting agents work on isolated branches while humans review and publish changes. Pipelines run on temporary branches and publish through a final merge, so all outputs become visible atomically or none do. Finally, we report production lessons as well as correctness insights from a preliminary Alloy model of our core abstractions.
\end{abstract}

\maketitle

\pagestyle{\vldbpagestyle}
\begingroup\small\noindent\raggedright\textbf{VLDB Workshop Reference Format:}\\
\vldbauthors. \vldbtitle. VLDB \vldbyear\ Workshop: \vldbworkshop.\\ 
\endgroup
\begingroup
\renewcommand\thefootnote{}\footnote{\noindent
This work is licensed under the Creative Commons BY-NC-ND 4.0 International License. Visit \url{https://creativecommons.org/licenses/by-nc-nd/4.0/} to view a copy of this license. For any use beyond those covered by this license, obtain permission by emailing \href{mailto:info@vldb.org}{info@vldb.org}. Copyright is held by the owner/author(s). Publication rights licensed to the VLDB Endowment. \\
\raggedright Proceedings of the VLDB Endowment. 
ISSN 2150-8097. \\
}\addtocounter{footnote}{-1}\endgroup

\ifdefempty{\vldbavailabilityurl}{}{
\vspace{.3cm}
\begingroup\small\noindent\raggedright\textbf{VLDB Workshop Artifact Availability:}\\
The source code, data, and/or other artifacts have been made available at \url{\vldbavailabilityurl}.
\endgroup
}

\section{Introduction}
\label{sec:intro}
Coding agents are taking the software engineering industry by storm, both when humans and agents write code together in a tight feedback loop and when agents in a ReAct loop \cite{yao2023reactsynergizingreasoningacting} continuously hill-climb toward a task. Version control systems such as \textit{Git} lie at the core of both patterns as they allow developers to incrementally develop software, using \textbf{commits} as intermediate checkpoints. Commits support time-travel for debugging and reverting code, and provide a unit of concurrent collaboration. However, agentic adoption in the data analytics domain lags behind the rest of the industry. The lakehouse is the \textit{de facto} standard OLAP for analytics and AI workloads~\cite{Tagliabue2023BuildingAS}; however, the affordances in traditional OLAP systems make agents unsafe~\cite{tagliabue2025trustworthyaiagenticlakehouse}.

We share the design of \texttt{GitLake}, the Git-for-data layer inside \texttt{Bauplan}'s lakehouse platform. As labor shifts from writing code to verifying and approving changes, correctness in the face of untrusted actors becomes non-negotiable \cite{liu_agentfirst_2025}. \texttt{GitLake} induces a natural division of labor between humans and agents: agents can explore and propose changes on isolated \textbf{branches}, while humans review and approve only what should be published to production. The core abstractions are obtained by ``porting'' \textit{Git} primitives to OLAP. By reusing familiar concepts, users quickly learn how to leverage the APIs to develop data pipelines collaboratively and time-travel to a previous state of the lake. We summarize our contributions as follows:

\begin{enumerate}
    \item we motivate Git-like abstractions in agentic data workflows through the lens of production workloads. As of today, \texttt{Bauplan} has run millions of jobs across hundreds of thousands of data branches (Section~\ref{sec:lessons}), making it, to our knowledge, one of the first systems of this kind tested at industry scale;
    \item we identify a core set of primitives (commits, branches, merges) building on top of single-table guarantees from open formats, and show that Git-for-data is a versatile mental model that can unify versioning, collaboration, and transactional guarantees, while also providing a natural human review boundary for agent-generated changes;
    \item we share lessons from our implementation journey (Section~\ref{sec:lessons}): copy-on-write storage, API design, and the use of lightweight formal modeling to stress-test core intuitions. We discuss real-world production metrics from agentic usage, and share with the community an open-source Alloy model.
\end{enumerate}

Notwithstanding our focus on safe agentic workloads on a production lakehouse, our design sits more generally at the intersection of frontier topics in data management and distributed systems. As such, we believe our lessons from the trenches to be valuable to a broad set of practitioners.

\section{Git: from code to data}
\label{sec:motivation}

While coding agents are powerful for code generation, they can rarely ``one-shot'' a complex or delicate task. We consider the ``eventual completion'' of such a task as a \textbf{development process}. \textit{Git} primitives are designed for modifying codebases that are built, tested, and run locally, and every change in \textit{Git} is immutable, but nothing is fatal. In particular, the interest in self-driving codebases highlights three key aspects in the development process:

\begin{description}
    \item[Fail safely] A multi-table (and often multi-language) data pipeline should not result in an inconsistent lakehouse state when it fails. \footnote{Industry lakehouses such as \textit{Snowflake} and \textit{Databricks} do not offer APIs for multi-language pipeline transactions.}
    \item[Cooperative work] A swarm of agents and a team of humans working in concert should be able to build on each other's work, in the spirit of branching off existing solutions, iterating, and managing conflicts through the tried-and-tested PR flow.
    \item[Backtrack] In the face of catastrophic failures, or when asked for an audit, the lakehouse should be returned to a previous sound state.
\end{description}

Comparable affordances in current OLAP systems are either rare or non-existent, leading to significant gaps. A pipeline that fails unsafely may result in an inconsistent global state where downstream readers observe a mix of old and new tables without any clear notion of lakehouse-wide success. Without an efficient and semantically sound workflow, data systems are left to constantly sync rows between development and production, with ad hoc reconciliation strategies. Additionally, in data, if AI-generated code drops a table, there may not be any obvious built-in way to undo the damage or provide point-in-time queries for auditing.

We could try to patch current systems to fill these gaps, for example, by swallowing the complexity of application-layer transactions to support manual pipeline rollbacks \cite{10.1145/3638553}. However, researchers are starting to question whether traditional platforms could ever deliver agentic data systems in a timely fashion \cite{liu_agentfirst_2025}. \texttt{GitLake} is built on the premise that \textit{Git}'s virtues are not accidental: our best bet is then to start treating our data estate as we treat our codebase.

\begin{figure}
\centerline{\includegraphics[width=0.45\textwidth]{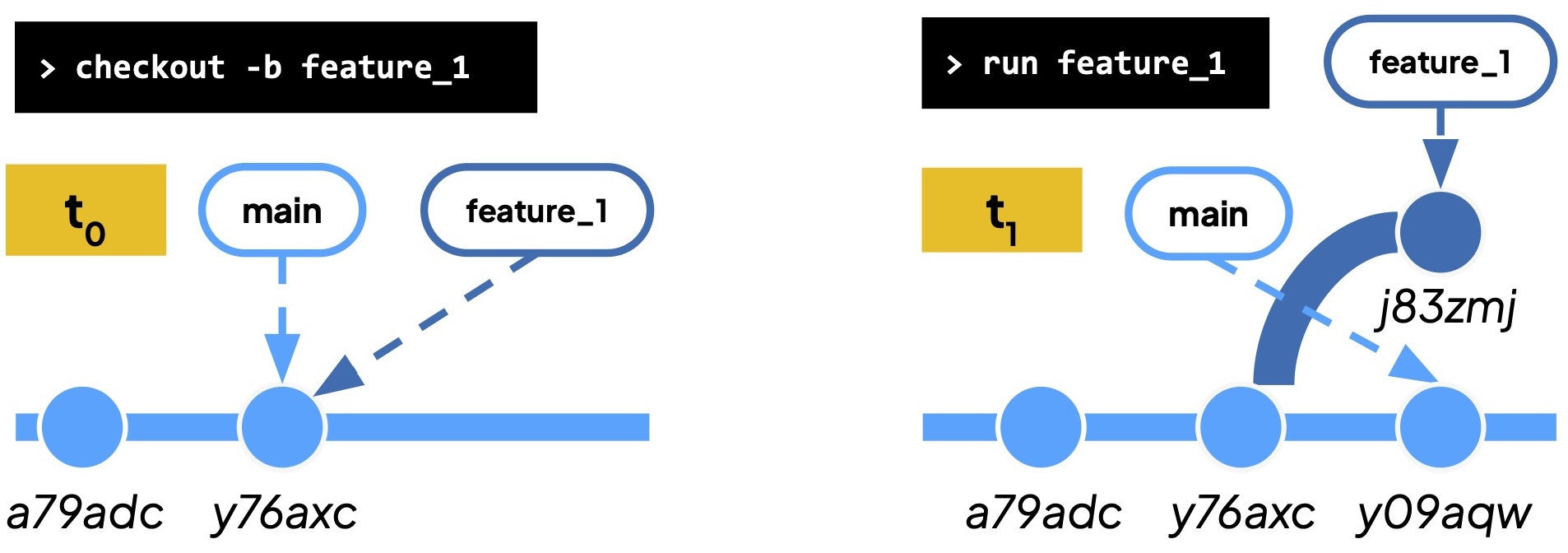}}
\caption{Branches are movable pointers: creating a new branch ($t_{0}$) is a no-op, but at $t_{1}$ \texttt{y76axc} is now the parent of two diverging commits, with pointers moving accordingly.}
\label{fig:branch}
\end{figure}

\section{System Design}
\label{sec:git}

Our design is motivated by the observation that Apache Iceberg already provides a strong primitive: ACID-compliant, \textit{single-table} snapshots, with optimistic locks guaranteed by a relational database at the catalog layer. We lift single-table writes to global lakehouse writes, and then show how to progressively build out more \textit{Git} abstractions. In Section~\ref{sec:implementation}, we discuss the implementation of this design through concrete APIs, workflows, and storage optimizations.

\subsection{From snapshots to commits}

A single Iceberg table evolves through snapshots, persisted in S3 and (importantly) also recorded in the Iceberg catalog (i.e., \textit{Postgres}). The key insight is to lift (within the atomic swap for the snapshot update) the table change into a \textbf{data commit} that maps, at that moment in time, all catalog tables to their snapshots. As we attach to every commit a hash identifier, metadata, and a parent pointer, we obtain our first \textit{Git} primitive in the data estate. It is already easy to see that, by adding a \texttt{hash} argument to a query API, we can provide time-travel for auditing and debugging with minimal changes to the read path: retrieve the relevant snapshots from the commit and pass the metadata URIs to the engine for scans. Since commits are stored in the catalog independently of the tables they reference, this implies that any destructive mutation (e.g. an agent dropping a table) is reversible through a revert API (Listing~\ref{lst:api}).

\subsection{From commits to branches}

By navigating from commit to commit through their parents, we naturally induce \textit{histories}: a \textbf{data branch} is simply a movable reference to the \textit{HEAD} of a history. When creating a branch \texttt{feature} from the production lakehouse (\texttt{main}), we initially get a new movable pointer to the same commit (Figure~\ref{fig:branch}): after writing to \texttt{feature}, \texttt{main} and \texttt{feature} diverge as \texttt{feature} now points to a new commit. It is then straightforward to see that, by adding a \texttt{reference} argument to a pipeline API, we can provide sandboxed development for data assets, as \texttt{run(..., ref=feature)} leaves downstream consumer queries intact.

\subsection{From branches to merges}

\begin{figure}
\centerline{\includegraphics[width=0.45\textwidth]{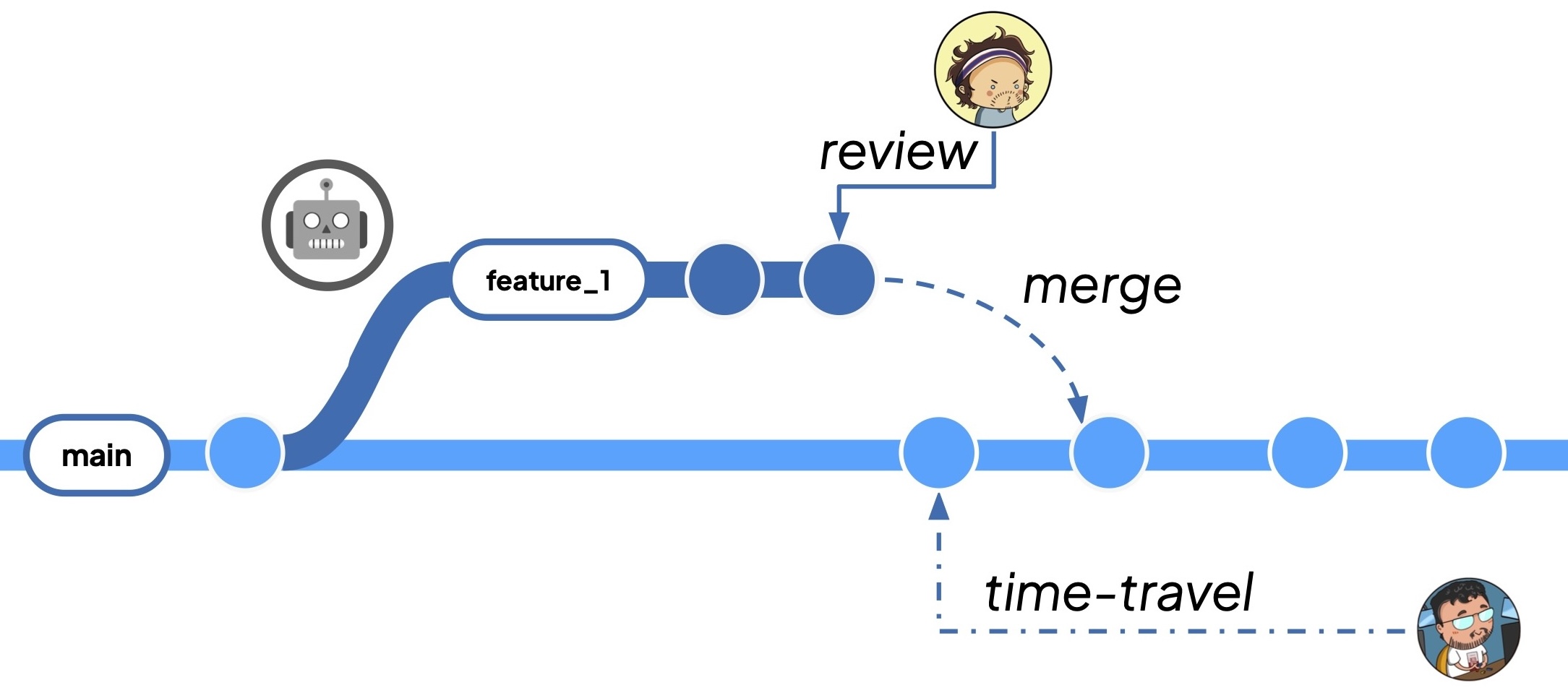}}
\caption{Git APIs enable collaboration and auditability.}
\label{fig:collab}
\end{figure}

A \textbf{data merge} takes two heads and produces a new commit on the destination branch that applies, pending conflicts, the snapshot updates reachable from the source and not yet present in the target. Crucially, merges happen \textit{atomically} and in the control plane only: merging is metadata-centric, i.e., a catalog update rather than moving or rewriting the underlying Parquet files, so collaboration remains cheap even when tables are large. By adding \texttt{merge}, collaborative scenarios discussed for codebases are enabled on data: Figure~\ref{fig:collab} depicts human reviews of agentic writes by leveraging the abstractions introduced so far. In practice, the merge into \texttt{main} also acts as a possible review boundary: agents can iterate freely on branches, but production publication happens only through an inspectable, human-approved merge. While collaboration may indeed be solved, this is not yet sufficient for correctness.

\subsection{From merges to transactions}
\label{sec:trans}

\begin{figure}
\centerline{\includegraphics[width=0.45\textwidth]{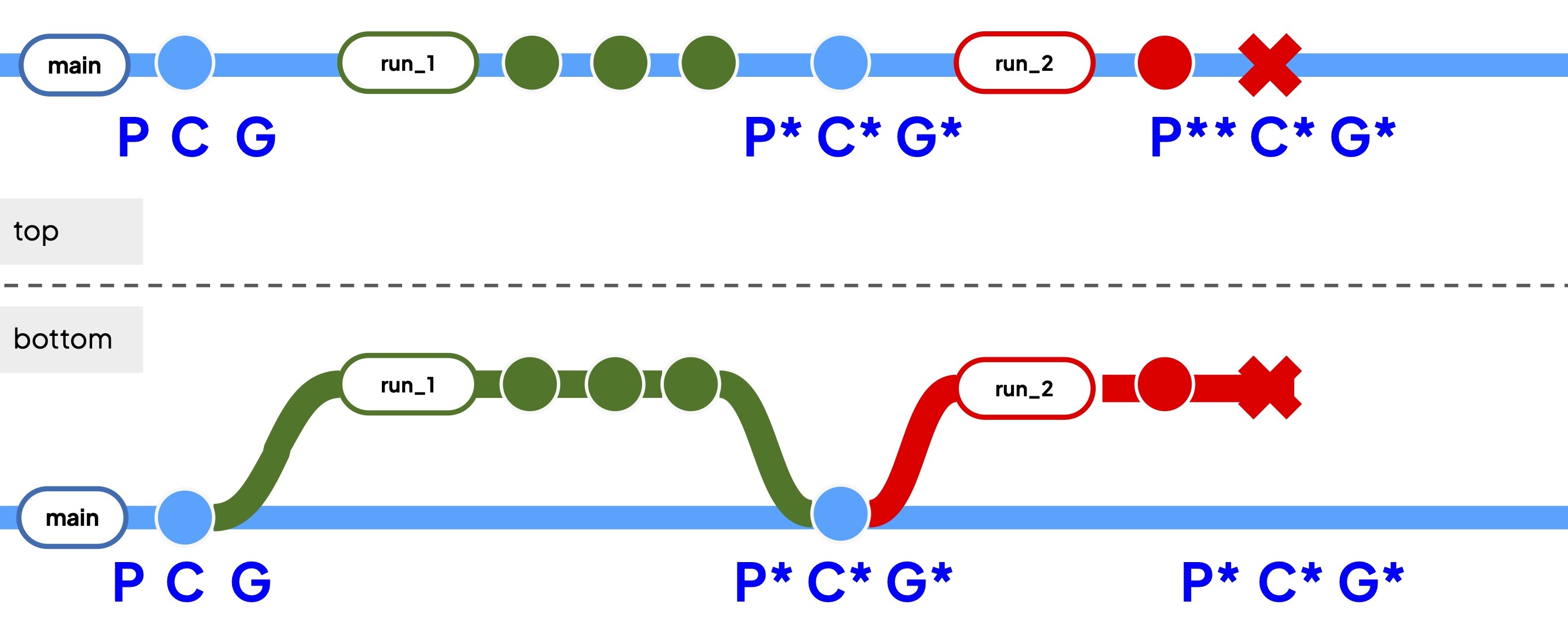}}
\caption{\textbf{Transactional pipelines}. \textit{Top}: without coupling temporary branches with runs, \texttt{run\_2} leaves \texttt{main} with a new version of \textit{Parent} but an old version of \textit{Child} and \textit{Grandchild}. \textit{Bottom}: the \texttt{run} API guarantees atomic publication of all tables on success, and isolation in case of failure.}
\label{fig:transaction}
\end{figure}

No primitive we have seen so far prevents the ``half-written pipeline'' (Section~\ref{sec:motivation}), as exemplified by Figure~\ref{fig:transaction} (top). \texttt{run\_1} executes successfully on \texttt{main}. Tables \textit{Parent}, \textit{Child}, and \textit{Grandchild}, abbreviated as \(P\), \(C\), and \(G\), are updated to snapshots \(P^{*}\), \(C^{*}\), and \(G^{*}\), respectively; however, \texttt{run\_2} breaks after updating \(P^{*}\) to \(P^{**}\) but before updating \(C^{*}\), leaving \texttt{main} in a \textit{globally} inconsistent state built out of legitimate single-table snapshots \(\{P^{**}, C^{*}, G^{*}\}\). Since \texttt{main} can be accessed at any point by downstream systems, the inconsistency may percolate uncontrollably.

Lakehouse pipelines run on ephemeral, multi-language compute and decoupled storage. To recreate MVCC-style transaction boundaries, we \textit{logically} bundle compute and \textit{merges} inside the \texttt{run()} API. Figure~\ref{fig:transaction} (bottom) illustrates \textbf{transactional branches}. First, \texttt{run\_1} shows the happy path: a temporary branch is opened to host commits generated by the pipeline, and it is merged at the end. When the merge happens, consumers see all the new snapshots at once. \texttt{run\_2} illustrates the unhappy path: failure to update \(C^{*}\) to \(C^{**}\) does \textit{not} compromise \texttt{main}, which continues to serve downstream consumers the globally consistent state of the first successful run.  As an additional bonus, the aborted transactional branch remains reachable for debugging, enabling users to triage the failure of \texttt{run\_2} by querying faulty intermediate assets.

\section{Implementation}
\label{sec:implementation}

\subsection{Data management}

The physical design of \texttt{GitLake} follows a common pattern in open lakehouse architectures: control state (such as branch heads, commit metadata, and run metadata) is mutable and stored in a relational catalog; data is immutable and stored using the Iceberg table structure (Parquet files and manifest files), simplifying bookkeeping and copy-on-write semantics. Appending data to \textit{Table T} in a new branch adds new Parquet files corresponding to the appended rows, avoiding costly duplication for pre-existing rows. In practice, most Git-for-data operations are metadata operations over references and do not involve data movement.

\subsection{APIs}

We expose the above primitives through APIs available through both the CLI and Python scripting. In line with agentic best practices, the CLI supports progressive discovery and self-documenting behavior through a recursive \texttt{--help} flag; the SDK exposes fully typed methods and supports local validation with type checkers for a fast feedback loop. Importantly, even subtle semantic distinctions are clearly marked through argument names and types; i.e., if a method accepts \texttt{branch=value} (and not \texttt{ref=value}), it is immediately clear that the semantics of the operation can only make sense at the \textit{HEAD} of a history. By implementing the CLI in Rust and then binding the same core methods into Python, we obtain two surfaces from a single source of truth that dispatch identical types and error taxonomies.

While a full description of the framework is beyond the scope of this paper \cite{sheng2026buildingcorrectbydesignlakehousedata}, Listing~\ref{lst:dag} shows a minimal DAG with two Python transformations chained together (\textit{Source} \(\rightarrow\) \textit{Parent} \(\rightarrow\) \textit{Child}). Listing~\ref{lst:api} then highlights how agents can programmatically control both the data assets and the transformations by interleaving data operations with Python control flow. Complex logic for creating, deleting, and merging branches can be assembled from simple typed primitives that are easy for agents to write and quick for humans to verify.

\begin{lstlisting}[
  language=Python,
  showstringspaces=false,
  columns=fullflexible,
  caption={A minimal DAG with two Python transformations.},
  label={lst:dag},
  basicstyle=\ttfamily\scriptsize,
  numbers=none
]
def parent_table(df: Source = source):
    # return a table fulfilling the Parent schema
    return table

def child_table(df: Parent = parent_table):
    # more transformation code here...
    return table
\end{lstlisting}

\begin{lstlisting}[
  language=Python,
  showstringspaces=false,
  columns=fullflexible,
  caption={Interleaving data-ops with Python control flow.},
  label={lst:api},
  basicstyle=\ttfamily\scriptsize,
  numbers=none
]
# 1) assuming we have a client, get the current head of main
cnt_main: Commit = client.get_commits(ref="main", limit=1)[0]
# 2) create a development branch from production
dev_br: Branch = client.create_branch("dev_br", from_ref="main")
# 3) run the DAG on the branch
run_state = client.run("pipeline/", ref=dev_br)
# 4) merge the branch into production on success
if run_state.success() and verification_passed():
    client.merge(dev_br, into="main")
    client.delete_branch(dev_br)
# 5) query the table as it was *before* the merge
rows = client.query("SELECT SUM(_S) FROM child", ref=cnt_main.hash)
# 6) revert a table to a previous snapshot
assert client.revert_table(
    table="child",
    source_ref=cnt_main.hash,
    into_branch="main",
)
\end{lstlisting}

Our key insight from Section~\ref{sec:trans} is to modify the semantics of running a pipeline and \textit{logically} couple function execution with data branches. A platform-level execution of a \texttt{run} simply implements, behind the scenes, the flow at the bottom of Figure~\ref{fig:transaction}: a branch is opened automatically from the target branch, the \textit{writes} are materialized there, and the branch is merged and deleted on success; on failure, \texttt{run\_2}'s transactional branch stays open, and \texttt{main} does not contain a partial state. Importantly, this optimization is only possible because of the declarative nature of both the SDK (i.e., agents only specify that a DAG should run on a branch, not how) and the framework (functions specify desired inputs and their schemas, not the physical I/O).

\section{Lessons learned}
\label{sec:lessons}

\begin{figure}
\centerline{\includegraphics[width=0.45\textwidth]{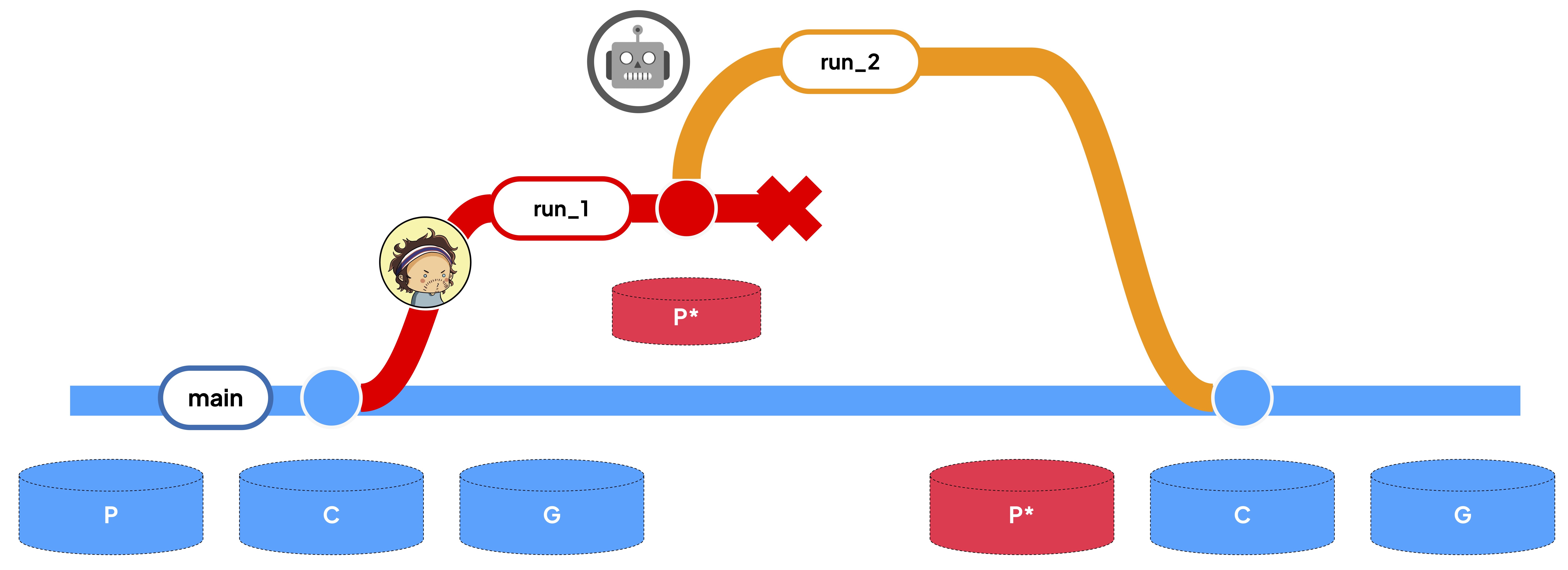}}
\caption{\textbf{A counterexample.} A failed run leaves an aborted branch open after the first commit. Another agent can branch off that commit and later merge back to \texttt{main}, creating an inconsistent state.}
\label{fig:counter}
\end{figure}

\makeatletter
\renewcommand\paragraph{\@startsection{paragraph}{4}{\z@}%
  {0.5ex \@plus .1ex \@minus .1ex}%
  {-0.9em}%
  {\normalfont\normalsize\itshape}}
\makeatother

\paragraph{Flexibility vs. correctness.} Adding \textit{Git} primitives to data DAGs expands the space of reachable states combinatorially. Inspired by the success of lightweight formal models in distributed systems \cite{Bornholt2021}, we ported our abstractions to Alloy to discover counterexamples that sharpen the intended semantics of the system: are inconsistent states really unrepresentable? Preliminary findings point to a tension between flexibility and correctness: Figure~\ref{fig:counter} illustrates a discovered counterexample to consistency in the face of failed runs on branches. Nested branches are powerful, so the obvious solution of disallowing branches on branches is not necessarily the right one. We leave further iterations to future work.

\paragraph{Branches grow quickly.} Organizations on \texttt{Bauplan} spawn branches at a high rate. In practice, we observe that our copy-on-write, metadata-only branching system scales to agentic usage and concurrency well beyond typical human-centric workloads. Traces from production confirm that creating a branch is effectively a no-op ($p_{95}$ is around 80ms), even at a pace of hundreds of thousands of new branches per week. As highlighted in our recent benchmarks 
\footnote{\url{https://github.com/BauplanLabs/OlapBranchBench}}, comparable primitives in Snowflake (zero-copy clone) and Databricks (shallow copy) are 100x slower than \texttt{GitLake}. Finally, since tables are generated from code (Listing~\ref{lst:dag}), merge conflicts happen only in the rare case of concurrent \textit{code} modification: in our traces, we see on average only ten conflicts per 100k attempts. 

\paragraph{Verification will soon be the bottleneck.}
As work shifts from writing to reviewing, a merge-centric worldview risks moving the bottleneck to a different layer: if every data analysis must be reviewed before merging, exploration scales faster than human verification. As a way forward, we have experimented with decoupling getting answers from reconciling a canonical table version: if we could query \textit{across} branches, users could trade ``partial'' answers for quicker response time \cite{tagliabue2026queryingoncesupervaluationismagentic}.

\paragraph{Autonomy needs a harness.} Git-for-data abstractions have already been shown to support Ralph-like scenarios such as self-healing data pipelines \cite{tagliabue2025safeuntrustedproofcarryingai}: because branches are cheap, agents can explore multiple strategies and a verifier can compare their outputs before a single strategy is merged into \texttt{main}. In our own experience, however, the nuances of the lakehouse are still hard for LLMs to fully internalize: for example, after \texttt{run\_2} (Figure~\ref{fig:transaction}) an agent could branch off \(P^{**}\), fix the error, and run the pipeline again \textit{from the second node}. By treating nested branches as ``durable execution'', agents could avoid re-computing everything at every trial. This observation led us to invest in purpose-built skills that complement model intelligence with specific operational knowledge \cite{schneider2026skillissuesdatacentricoptimization}.

\section{Related work}

The original semantics for \textit{Git} are given in \cite{10.1145/2661136.2661137}: while code and data share similarities, table-backed primitives are novel. Our work aligns with database literature highlighting similarities between \textit{Git} primitives and transactions. Dolt~\cite{dolt} is an OLTP-focused branching database, lacking a lakehouse-oriented merge pattern. A \textit{Git} graph was modeled as a form of transaction in~\cite{yilmazdittrich}, but we focus on a different correctness boundary, as the dominant failure mode is partial publication rather than tuple-level anomalies. Nessie~\cite{nessie} provides Git-like versioning for lakehouse tables. \texttt{GitLake} differs in two ways: it optimizes common branch/catalog operations for high-frequency agentic workloads (up to 25x faster in standard CRUD-like requests), and it integrates versioning with pipeline execution so that the \texttt{run} API provides atomic publication across multi-table DAGs.

\section{Conclusion}

We described \texttt{GitLake}, the Git-for-data abstractions powering an agentic lakehouse. By lifting single-table evolution to commits, branches, merges, and reverts, we obtain a compact programming model for collaboration, reproducibility, rollback, and transactions. As of today, \texttt{Bauplan} has run millions of jobs across hundreds of thousands of data branches: the lessons we shared suggest that scaling systems to agentic scale will require rethinking most of the data stack.

\bibliographystyle{ACM-Reference-Format}
\bibliography{sample}

\end{document}